\documentclass[letterpaper]{article} 
\usepackage{aaai2026}  
\usepackage{times}  
\usepackage{helvet}  
\usepackage{courier}  
\usepackage[hyphens]{url}  
\usepackage{graphicx} 
\usepackage{natbib}  
\usepackage{caption} 
\usepackage{algorithm}
\usepackage{subcaption}
\usepackage{booktabs}
\usepackage{tabularx} 
\usepackage{algpseudocode}
\usepackage{amsmath}
\usepackage{adjustbox} 
\usepackage{times}
\usepackage{courier}
\usepackage{xcolor}
\usepackage{hyperref}
\usepackage{newfloat}
\usepackage{listings}
\DeclareCaptionStyle{ruled}{labelfont=normalfont,labelsep=colon,strut=off} 
\floatstyle{ruled}
\newfloat{listing}{tb}{lst}{}
\floatname{listing}{Listing}
\title{ARISE: Agentic Rubric-Guided Iterative Survey Engine\\
for Automated Scholarly Paper Generation}

\author{
  Zi Wang\textsuperscript{\rm 1},
  Xingqiao Wang\textsuperscript{\rm 1},
  Sangah Lee\textsuperscript{\rm 2},
  Xiaowei Xu\textsuperscript{\rm 1}\thanks{Corresponding author.}
}
\affiliations{
  \textsuperscript{\rm 1} University of Arkansas at Little Rock\\
  \textsuperscript{\rm 2}Seoul National University\\
  \texttt{zwang@ualr.edu, xwang1@ualr.edu, sanalee@snu.ac.kr, xwxu@ualr.edu}
}

\nocopyright
\begin{document}

\maketitle

\begin{abstract}
The rapid expansion of scholarly literature presents significant challenges in synthesizing comprehensive, high-quality academic surveys. Recent advancements in agentic systems offer considerable promise for automating tasks that traditionally require human expertise, including literature review, synthesis, and iterative refinement. However, existing automated survey-generation solutions often suffer from inadequate quality control, poor formatting, and limited adaptability to iterative feedback—core elements intrinsic to scholarly writing.

To address these limitations, we introduce ARISE, an Agentic Rubric-guided Iterative Survey Engine designed for automated generation and continuous refinement of academic survey papers. ARISE employs a modular architecture composed of specialized large language model agents, each mirroring distinct scholarly roles, such as topic expansion, citation curation, literature summarization, manuscript drafting, and peer-review-based evaluation. Central to ARISE is a rubric-guided iterative refinement loop where multiple reviewer agents independently assess manuscript drafts using a structured, behaviorally anchored rubric, systematically enhancing the content through synthesized feedback.

Evaluating ARISE against state-of-the-art automated systems and recent human-written surveys, our experimental results demonstrate superior performance, achieving an average rubric-aligned quality score of 92.48. ARISE consistently surpasses baseline methods across metrics of comprehensiveness, accuracy, formatting, and overall scholarly rigor. All code, evaluation rubrics, and generated outputs are provided openly at
\href{https://github.com/ziwang11112/ARISE}{\textcolor{blue}{https://github.com/ziwang11112/ARISE}}.

\end{abstract}

\section{Introduction}
\begin{figure*}[t]
  \centering
  \includegraphics[width=\textwidth]{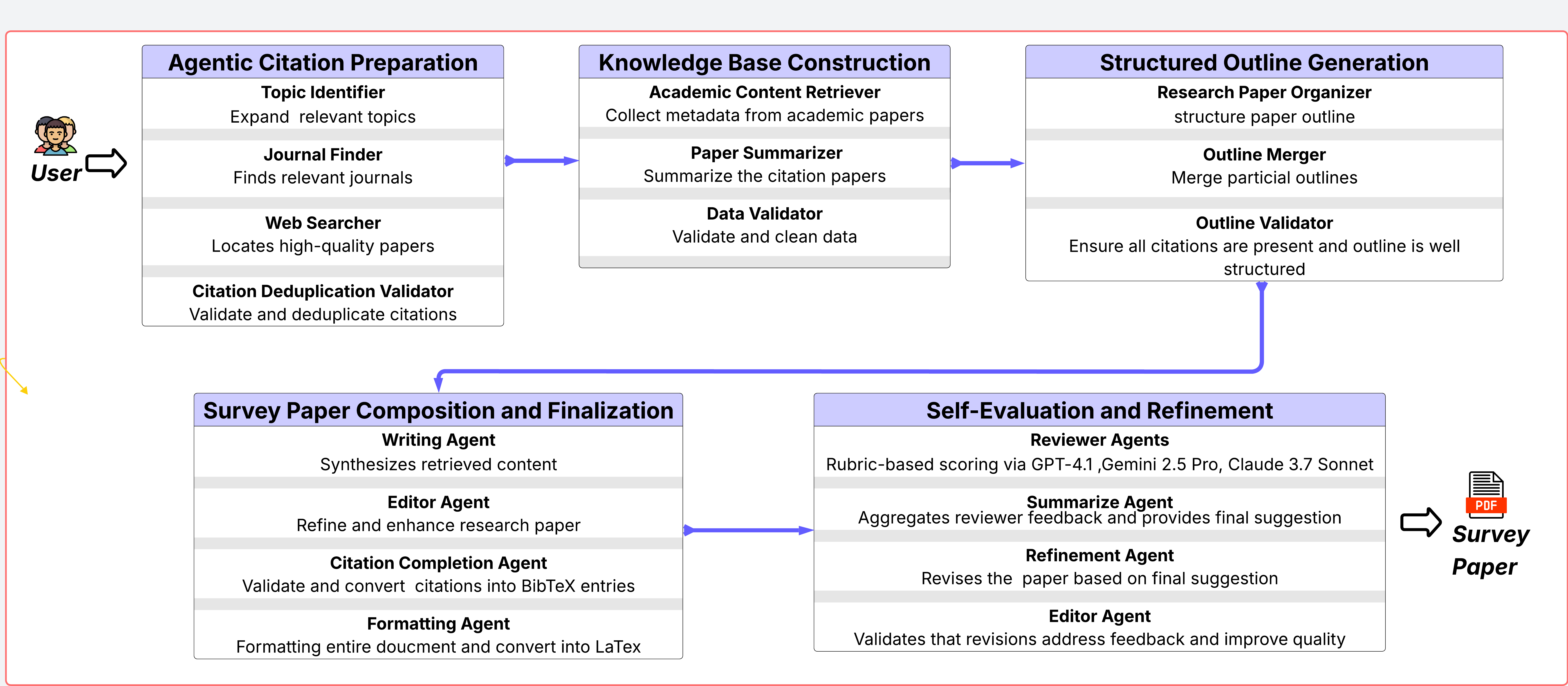}
  \caption{Agent roles and functionality in ARISE. Each agent is assigned a specific function in the modular survey generation and refinement pipeline. All non-reviewer agents run on GPT-4.1 by default. Reviewer agents use a cross-family judge pool (GPT-4.1(optional), Gemini 2.5 Pro, and Claude 3.7 Sonnet).}
  \label{fig:overview}
\end{figure*}

Recent advances in agentic AI have demonstrated the potential of large language model (LLM) agents to collaborate, reason, and solve complex tasks by mirroring human workflows. By orchestrating multiple specialized agents in a modular, feedback-driven manner, agentic systems offer a powerful paradigm for automating labor-intensive processes that traditionally require expert human collaboration, and have already proven effective in domains such as code generation, multi-step planning, and academic research systems.

Survey paper writing is a promising application for agentic systems, as a subdomain of academic research systems, it demands coordinated, expert-level reasoning and reflects the complexity of real-world scholarly workflows. With scientific literature expanding rapidly~\cite{ref106}, synthesizing new developments into high-quality surveys has become increasingly challenging. Recent studies have explored automating survey paper generation with LLMs, combining retrieval, structuring, and generation in end-to-end pipelines. Notable approaches such as \textit{AutoSurvey}\cite{ref10}, \textit{SurveyX}\cite{ref11}, and \textit{SurveyForge}~\cite{ref12} demonstrate promising results by leveraging LLM-assisted writing and hybrid retrieval techniques. 

While recent literature review automation methods offer promising performance ~\cite{Wu2025}, they still have notable limitations. For example, these approaches commonly rely on preprint-heavy sources, generate survey paper within one pass running and lacking efficient evaluation standards. As a result, ensuring high-quality, reliable references is challenging, as retrieval agents often surface outdated, non-peer-reviewed, or low-quality sources. In addition, real-world survey writing is inherently iterative, requiring multiple cycles of review and revision—a process not well captured by existing single-pass systems. Finally, although benchmarking and evaluation protocols are improving, comprehensive rubrics and peer-review-style feedback remain rare, limiting the interpretability, transparency, and reproducibility of automated outputs.

To address these limitations, we propose \textbf{ARISE}, an agentic system that decomposes the academic survey writing process into specialized LLM-powered agents, each mirroring a distinct human role. It employs a \emph{citation-first, article-level} retrieval pipeline that prioritizes references from reputable, peer-reviewed journals and conferences, and produces fully editable \LaTeX{} manuscripts with structured bibliographies tailored to target publication venues. Dedicated validation agents at every stage ensure factual accuracy and minimize hallucinations. ARISE further incorporates a rubric-guided, multi-agent iterative refinement loop: powerful LLMs serve as distinct reviewer agents, each applying a shared evaluation rubric to assess system outputs and generate structured feedback. This feedback is synthesized by the refinement module, driving systematic and interpretable improvements through multiple revision cycles while ensuring transparency and reproducibility.

\noindent\textbf{Our main contributions are:}
\begin{itemize}
     \item We present \textbf{ARISE}, an agentic system that automates end-to-end survey generation and peer-review, supporting full-cycle scholarly workflows with modular, specialized LLM agents.
    \item We introduce a citation-first, rubric-guided, multi-agent iterative refinement framework that produces template flexible outputs, enabling systematic and transparent quality improvement through structured reviewer feedback.   
    \item We introduce an extensible, behaviorally anchored rubric and evaluation framework that mirrors human peer review, enabling transparent, reproducible, and customizable assessment.
\end{itemize}

\section{Related Work}
\label{sec:related_work}

\subsection{Agentic System Collaborative Frameworks}

Recent advances in large language models have spurred the development of frameworks for orchestrating agentic systems, including CrewAI~\cite{crewai2024}, AutoGen~\cite{wu2023autogen}, and LangGraph~\cite{langgraph2024}. These toolkits provide abstractions for designing and coordinating teams of LLM-powered agents in hierarchical, collaborative, and conversational workflows. 

Further progress includes frameworks for emergent behavior and specialized capabilities, such as HuggingGPT~\cite{shen2023hugginggpt}, MM-Agent~\cite{yang2023mmagent}, AgentVerse~\cite{li2024agentverse}, and Any-Agent~\cite{shen2023anyagent}, enabling dynamic tool use, web-based interaction, and multi-modal collaboration. Additional research explores communicative agent systems for software engineering~\cite{qian2023communicativeagents,he2025llmbasedmultiagentsystemssoftware}, collective reasoning~\cite{du2023multiagent}, and open challenges in cooperative AI~\cite{dafoe2021open}, as well as scientific discovery~\cite{hart2023autonomous}.

These advances have made it increasingly feasible to design, prototype, and benchmark multi-agent LLM systems for tasks such as collaborative writing, complex reasoning, tool use, and structured document generation. Our work builds on this foundation, leveraging agentic orchestration and modular task decomposition for robust and extensible academic survey generation.

\subsection{LLM-Based Survey Generation}

Recent systems have explored automating survey paper generation with large language models (LLMs), combining retrieval, structuring, and generation in end-to-end pipelines. \textit{AutoSurvey}~\cite{ref10} follows a four-phase pipeline with arXiv-based retrieval and LLM-assisted writing, but relies solely on preprints and offers limited source curation. \textit{SurveyX}~\cite{ref11} enhances retrieval via hybrid keyword expansion and semantic filtering with an \textit{AttributeTree} citation structure, though it could be improved in modularity and user control. \textit{SurveyForge}~\cite{ref12} uses heuristic templates and a memory-driven Scholar Navigation Agent (SANA), and introduces the SurveyBench benchmark for holistic evaluation, but remains constrained by a static architecture and preprint-heavy sourcing.

\subsection{LLM Evaluation and Rubric Design}

Evaluation of LLM-generated outputs remains a major challenge~\cite{chang2023surveyevaluationlargelanguage,guo2023evaluatinglargelanguagemodels,laskar2024}, especially for complex tasks like survey writing, due to subjectivity, inconsistency, and ambiguity in assessment criteria. Several recent works have addressed these issues by proposing more comprehensive, behaviorally anchored rubrics and checklists~\cite{10.1145/3636515, lee2024checkeval}. Peer-review guidelines from organizations such as IEEE and ACL~\cite{ieee-reviewer-guidelines, aclrr2025} provide additional best practices for rubric construction and reviewer alignment.

\section{Methodology}
\label{sec:methodology}
\subsection{System Overview}

Figure~\ref{fig:overview} presents the modular architecture of \textsc{ARISE}, designed for automated survey generation and iterative self-improvement. We target a systematic mapping review with article‑level screening and structured rubric‑guided synthesis. Each module consists of one or more dedicated LLM agents, assigned to perform specific scholarly tasks such as topic expansion, citation discovery and validation, literature summarization, outline drafting, manuscript generation, and peer review. In total, \textsc{ARISE} orchestrates 22 specialized agents, including subagents for title and abstract generation, section-level summarization, and citation completion. Some roles (e.g., “Summarizer”) are instantiated with module-specific prompts and objectives, reflecting the unique requirements of each pipeline stage. A complete roster of agents, along with their task descriptions and configurations, is provided in our supplementary material to support transparency and reproducibility.

\subsection{Citation Preparation}
\label{sec:phase1_citation}

\begin{figure}[t]
\centering
\includegraphics[width=0.95\linewidth]{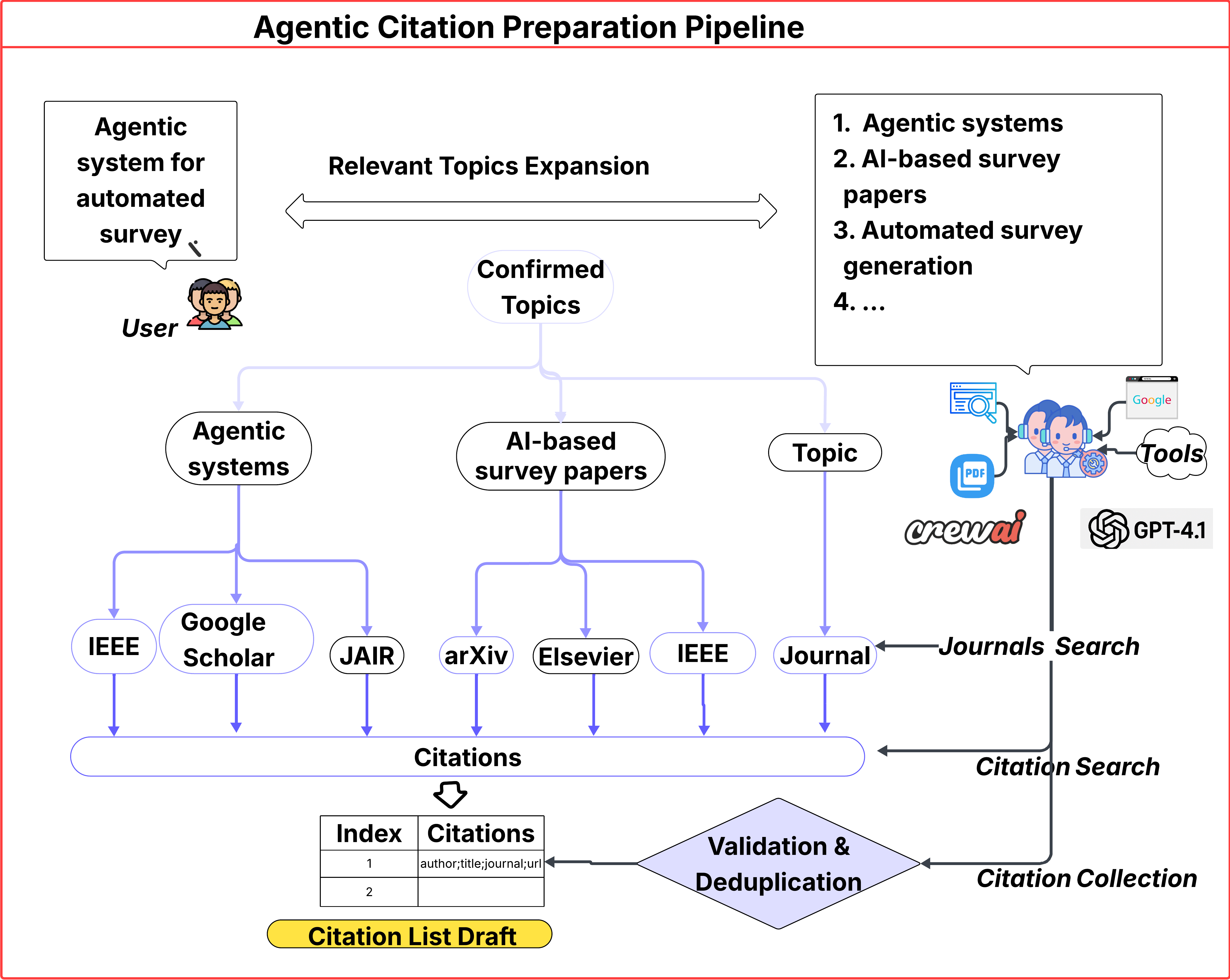}
\caption{Agentic citation preparation pipeline. Users shape topics and guide domain scoping; agents handle retrieval, filtering, and validation.}
\label{fig:phase1_pipeline}
\end{figure}

Figure~\ref{fig:phase1_pipeline} illustrates the agentic citation preparation pipeline in \textsc{ARISE}. The process begins with the user specifying an initial survey theme (e.g., “agentic systems for automated survey generation”). An expansion agent then proposes semantically related subtopics, which the user can further refine or approve.This interactive expansion step determines the conceptual scope of the survey and ensures both thematic breadth and alignment with user goals.

Once topics are confirmed, a \emph{domain-scoping} agent suggests publication venues that are \emph{topically appropriate} for each subtopic, including publisher portals (e.g., IEEE Xplore, Elsevier/ScienceDirect), academic search/indexing services (e.g., Google Scholar, Crossref, Semantic Scholar), and open-access repositories (e.g., arXiv). The goal of this step is to avoid obviously unrelated domains (e.g., business or biology journals when the topic is AI/ML), not to use venue prestige as a proxy for article quality. In other words, venues act as \emph{filters on field}, while inclusion and ranking decisions remain at the \emph{article level}.

For each (topic, source) pair, a citation retrieval agent gathers candidate references enriched with metadata such as authors, title, venue, year, and URL. Candidates retrieved from different sources are unified and normalized, then passed through automatic de-duplication and formatting validation to remove near-duplicates and malformed entries. The resulting curated citation list forms the basis for downstream knowledge base construction and outline generation.

\subsection{Structured Knowledge Base Construction}
\label{sec:phase2_kb}

\begin{figure}[t]
  \centering
  \includegraphics[width=\linewidth]{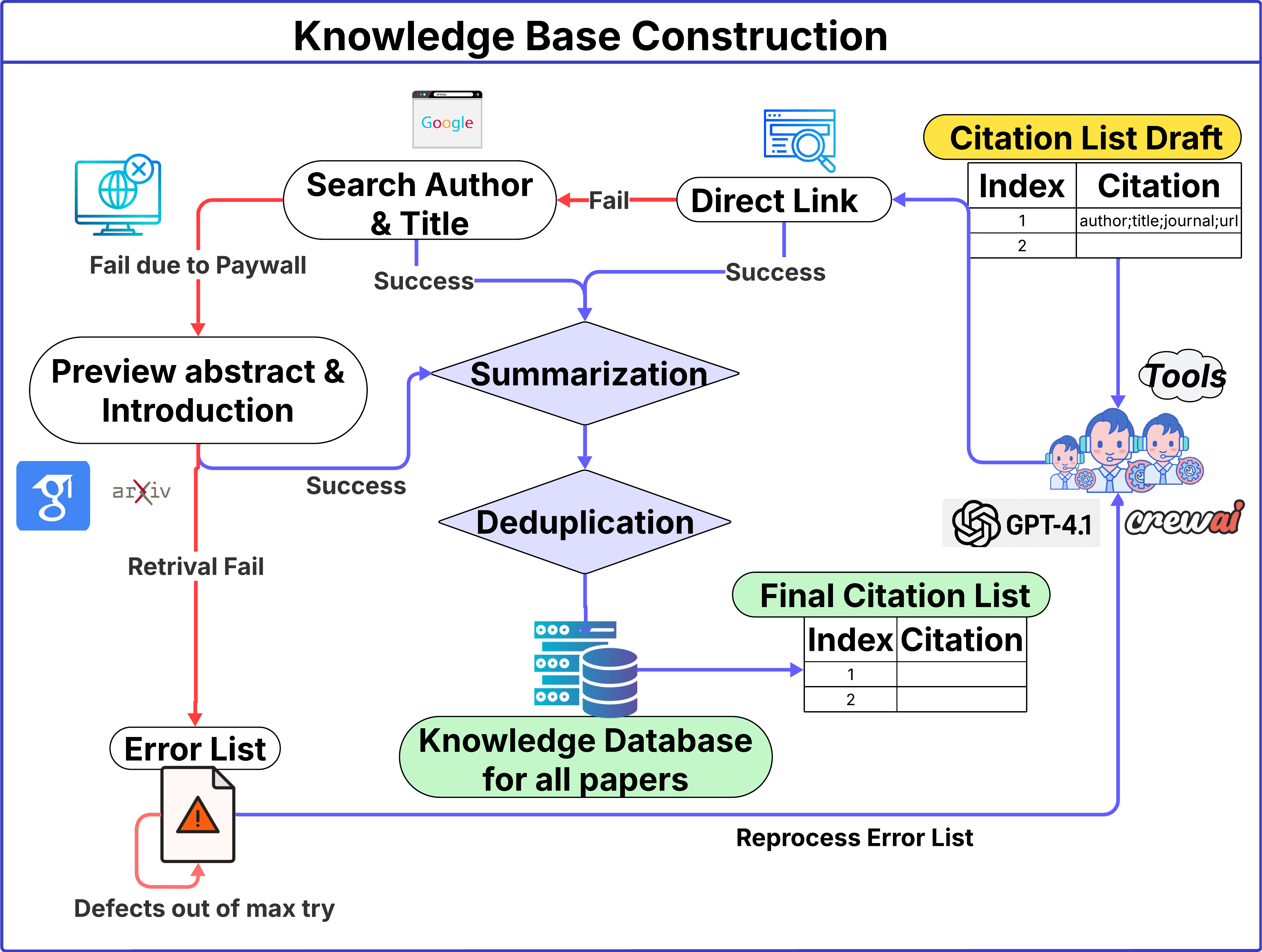}
  \caption{Structured Knowledge Base Construction. Each citation is processed via direct or fallback retrieval, summarized, deduplicated, and indexed in a persistent database aligned with the citation index.}
  \label{fig:phase2}
\end{figure}

Figure~\ref{fig:phase2} presents the structured knowledge base construction pipeline in \textsc{ARISE}. The process begins with the curated citation list from the previous phase or user-provided references. For each citation, the system first attempts direct retrieval via stored URLs. If a direct link fails due to paywalls or missing resources, a fallback metadata search is performed using author and title information on platforms such as Google Scholar or arXiv.

If full or partial content (such as the abstract or introduction) is successfully retrieved, an agent summarizes the text into a concise, contribution-focused entry. If all retrieval attempts fail, the citation is recorded in an \emph{Error List} for later review and possible reprocessing.

All retrieved summaries are then deduplicated and validated before being indexed into a persistent knowledge base, organized by citation index \texttt{refN}. This knowledge base forms the factual backbone for downstream modules, ensuring that the subsequent outline and paper composition phases are grounded in coherent, context-rich information.

\paragraph{Citation-Keyed Memory (CKM).}
Beyond serving as a passive repository, the knowledge base functions as a \emph{citation-keyed memory}: each entry is stored as a key–value pair $(\texttt{refN} \rightarrow \text{summary})$. During drafting and refinement, sections query CKM only with the citation keys they already use (i.e., $\mathrm{cite}(S)$ for a section $S$), and the system injects just those summaries into the prompt. This design minimizes irrelevant context and interference while preserving traceability from generated text back to its evidence sources.

\subsection{Structured Outline Generation}
\label{sec:phase3_outline}

\begin{figure}[t]
  \centering
  \includegraphics[width=\linewidth]{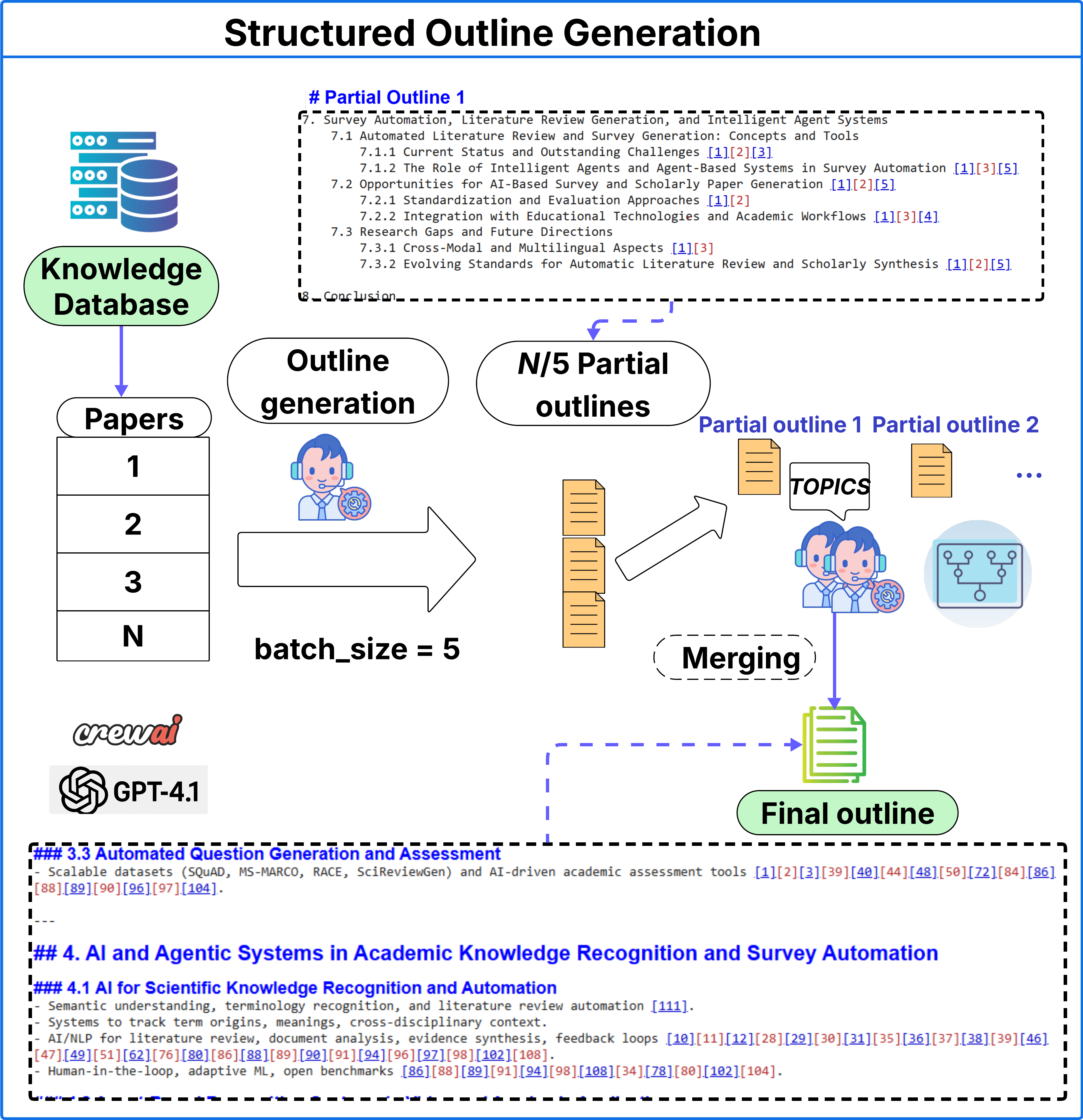}
  \caption{Structured outline generation component. Citation summaries are grouped, outlined, and iteratively merged to form a thematically coherent, citation-preserving global structure.}
  \label{fig:phase3}
\end{figure}

To support coherent and citation-grounded survey writing, the system uses a team of agents to build a structured outline based on the knowledge base. As shown in Figure~\ref{fig:phase3}, the process starts from a knowledge base containing $N$ cleaned citation summaries. These are divided into mini-batches, and each batch is processed by a Writing Agent, which generates a partial outline with sections, subsections, and explicit citation index references (e.g., [1][2][3]).

The partial outlines are then passed to a Merging Agent, which combines them in pairs to form progressively larger outlines. After each merge, a Validation Agent checks that the merged outline is coherent and well organized, and that no newly introduced redundancies or obvious gaps appear.

\paragraph{Citation-Preserving Outline Synthesis (CPOS).}
Beyond semantic coherence, we enforce a \emph{citation-preserving invariant} during merging: for any pair of outlines $A$ and $B$ with citation index sets $\mathrm{cite}(A)$ and $\mathrm{cite}(B)$, the merged outline $C$ must satisfy
$\mathrm{cite}(C) = \mathrm{cite}(A) \cup \mathrm{cite}(B)$.
After each merge, the Validation Agent compares citation sets and identifies any missing indices; a backfilling step then re-inserts the corresponding references and local context. This merging and validation cycle continues until a single, citation-complete Final Outline is created, and a final check ensures that all original references from the curated citation index are present.

\subsection{Survey Paper Composition and Finalization}
\label{sec:survey_composition}

This component completes the initial drafting and formatting stages of the system. It transforms the structured outline and curated knowledge base into a citation-grounded, academically formatted survey.

\begin{figure}[t]
  \centering
  \includegraphics[width=0.95\linewidth]{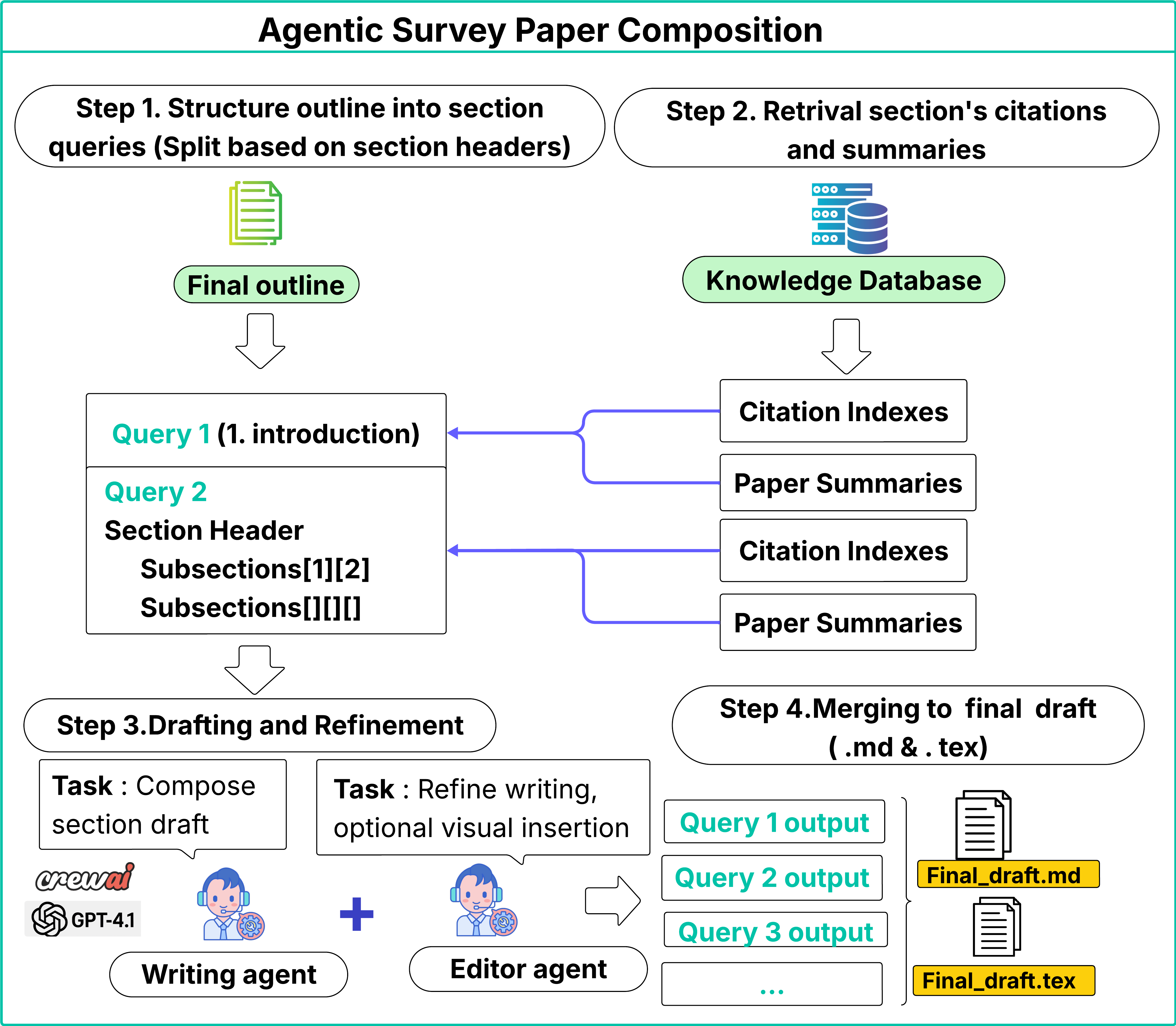}
  \caption{Agentic Survey Paper Composition. Each section query is matched with relevant citations and summaries. A writing agent drafts the content, which is refined by an editor agent before merging into intermediate content outputs.}
  \label{fig:phase4}
\end{figure}

As shown in Figure~\ref{fig:phase4}, the process begins by decomposing the final outline into section-level prompts based on the document hierarchy established during outline generation. For each section $S$, the system retrieves the relevant citation indices $\mathrm{cite}(S)$ and their associated summaries from the citation-keyed memory (CKM). A Writing Agent then synthesizes these section-specific summaries into coherent, thematically organized prose, ensuring that every claim is anchored in the cited works and that traceability to original sources is preserved. The resulting drafts are passed to an Editor Agent, which improves logical flow and clarity, resolves local redundancies, and inserts placeholders for tables or figures where appropriate. These refined sections are finally merged into a unified draft that respects the outline structure and maintains citation alignment.

\begin{figure}[t]
  \centering
  \includegraphics[width=0.95\linewidth]{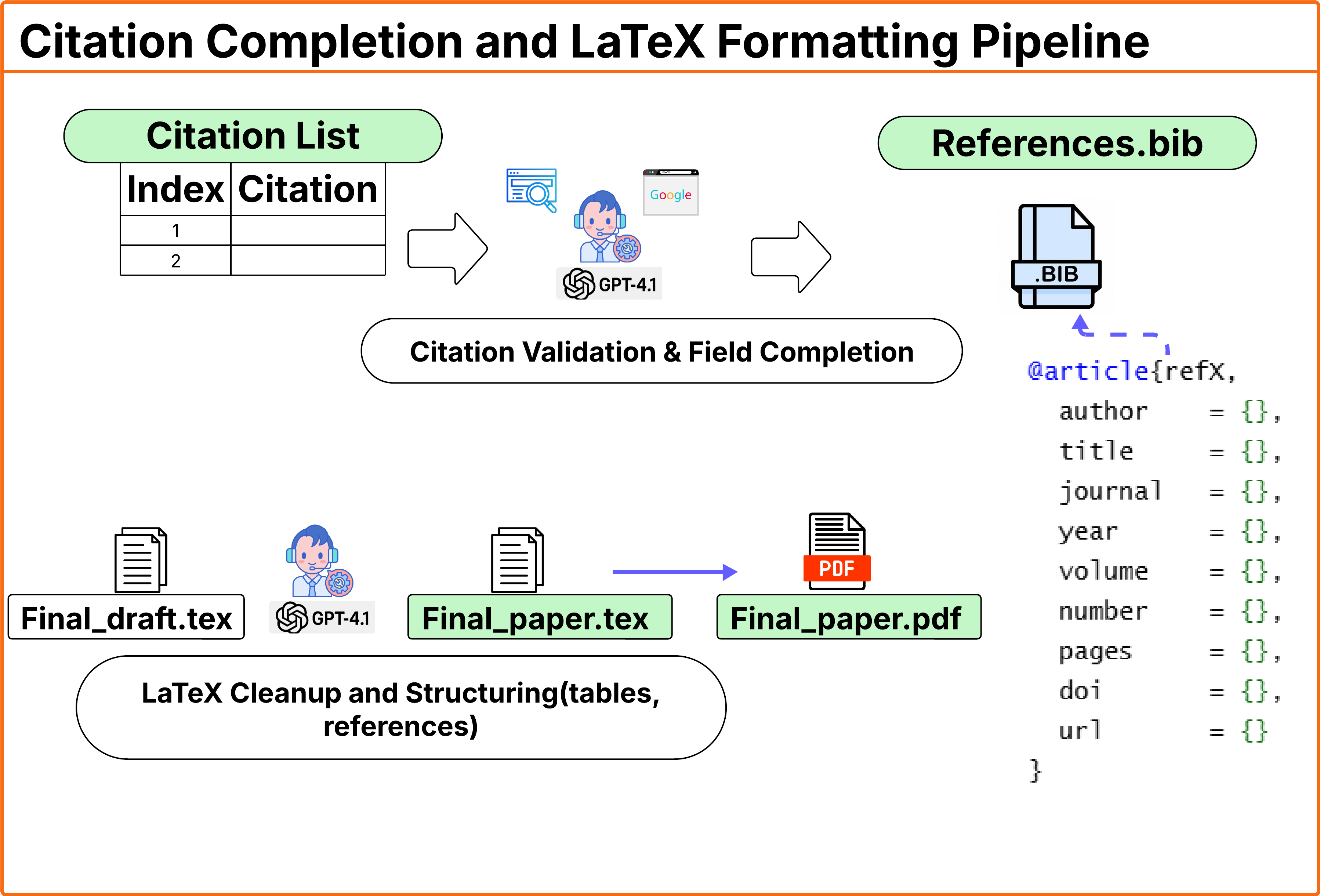}
  \caption{Citation and Formatting Pipeline. The system completes citation metadata, generates BibTeX entries, and standardizes the \LaTeX{} document to produce a clean, structured PDF ready for academic use.}
  \label{fig:phase5}
\end{figure}

\paragraph{Citation and Formatting Hygiene.}
Figure~\ref{fig:phase5} illustrates the citation completion and \LaTeX{} formatting pipeline. Starting from a citation list that may contain incomplete bibliographic information, a Citation Completion Agent validates each entry and resolves missing metadata fields (e.g., DOIs, venues, years) by querying trusted academic databases and search engines. The agent emits standardized Bib\TeX{} entries, which are compiled into a structured bibliography.

A Formatting Agent then enforces \emph{hygiene normalization} on the \LaTeX{} manuscript: it standardizes citation commands (e.g., transforming raw numeric brackets into consistent \verb|\cite{refN}| calls), harmonizes section and table environments, and removes residual artifacts from earlier stages of the pipeline. To ensure professional presentation and reliable compilation, the agent applies structured table environments and consistent style conventions throughout the document. The cleaned \LaTeX{} source can then be compiled into a camera-ready PDF under the target conference or journal template.

\subsection{Agentic Rubric-Guided Iterative Refinement Framework}

\begin{figure}[t]
  \centering
  \includegraphics[width=0.95\linewidth]{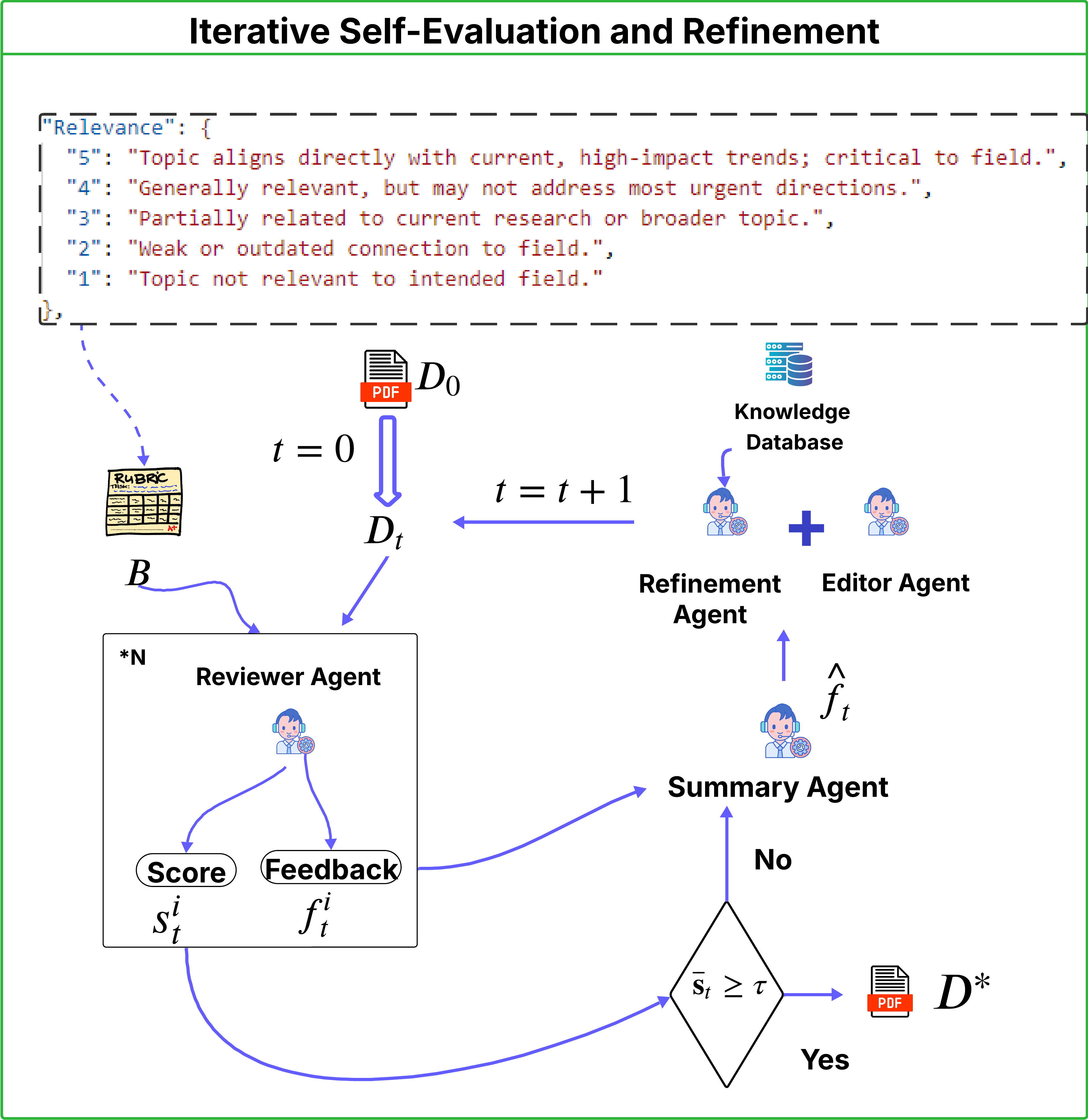}
  \caption{
  Agentic rubric-guided iterative refinement. At each iteration $t$, reviewer agents independently score draft $D_t$ using rubric $\mathcal{B}$, producing scores $s_t^i$ and feedback $f_t^i$. The average score $\overline{s}_t$ is computed; if $\overline{s}_t \geq \tau$, the draft is accepted as $D^*$. Otherwise, feedback is synthesized into a revision plan that drives targeted updates to produce $D_{t+1}$. The process repeats until the threshold is met or a stopping condition is reached.
  }
  \label{fig:phase6}
\end{figure}

\paragraph{Framework Overview.}
To mirror real-world peer review, ARISE treats refinement as a \emph{multi-agent control loop}. At each iteration \( t \), the current draft \( D_t \) is segmented into contiguous page chunks and independently evaluated by a set of reviewer agents \( \mathcal{R} \), each applying the shared rubric $\mathcal{B}$. Reviewer $i$ returns a rubric-based score \( s_t^{i} \) and structured feedback \( f_t^{i} \). Scores are then aggregated to compute an average quality estimate:
\begin{equation}
\overline{s}_t = \frac{1}{|\mathcal{R}|} \sum_{i \in \mathcal{R}} s_t^{i},
\label{eq:average_score}
\end{equation}
where \( |\mathcal{R}| = N \) is the number of reviewer agents.

If the aggregated score \(\overline{s}_t\) meets or exceeds the target threshold \(\tau\), the draft is accepted as the final output \( D^* \). Otherwise, a summary agent plays the role of a meta-reviewer: it synthesizes the individual feedback items $\{f_t^{i}\}$ into an actionable revision plan \( \hat{f}_t \). This plan specifies \emph{which sections and issues} to address (e.g., missing related work, weak analysis, unclear structure). A refinement agent and an editor agent then apply the plan to obtain an improved draft \( D_{t+1} \), and the loop repeats until the threshold or another stopping condition (e.g., max rounds) is reached. Appendix~1 shows concrete examples of reviewer feedback and the resulting revision plans.

\paragraph{Evidence-Locked Targeted Revision.}
Crucially, ARISE couples this control loop with \emph{evidence-locked, section-level revision}. For a given section $S$, we first recover its set of cited keys $\mathrm{cite}(S)$ (i.e., the \texttt{refN} indices that appear in the text) and query the citation-keyed memory (CKM) to obtain only the corresponding summaries $(\texttt{refN} \rightarrow \text{summary})$. The refinement agent is instructed to:
\begin{itemize}
  \item modify \emph{only} the sections that are explicitly targeted in $\hat{f}_t$ (no global free-form rewriting), and
  \item ground new wording exclusively in the CKM entries for $\mathrm{cite}(S)$, without introducing new, uncited claims or references.
\end{itemize}
This evidence-locked sectional refinement (ELSR) minimizes hallucination and scope drift by design, while preserving traceability from every revision back to the underlying sources.

\paragraph{Bias Controls and Early Stopping.}
To reduce LLM-as-judge bias, we use a cross-family reviewer pool and report both a \emph{tri-judge} setting (GPT-4.1, Gemini 2.5 Pro, Claude 3.7 Sonnet) and a setting where the generator family is excluded from the judge set. We also vary the number of reviewers $N$, the number of refinement rounds $t$, and the acceptance threshold $\tau$ to probe quality--cost trade-offs. In all cases, we log chunk-level scores and aggregated trajectories $\{\overline{s}_t\}_{t=0}^T$, enabling analysis of convergence behavior and diminishing returns.

\paragraph{Rubric Construction and Evaluation Rationale.}
Our shared rubric $\mathcal{B}$ is constructed by synthesizing best practices from established peer-review guidelines—specifically, the IEEE~\cite{ieee-reviewer-guidelines} and ACL Rolling Review~\cite{aclrr2025}—as well as recent advances in automated, rubric-based evaluation~\cite{10.1145/3636515}. We further incorporate insights from frameworks such as CheckEval~\cite{lee2024checkeval}, which emphasize subdividing evaluation criteria into explicit, behaviorally anchored sub-aspects to improve reliability and inter-rater agreement in LLM-based assessment.

To ensure consistent and interpretable scoring, $\mathcal{B}$ operationalizes seven core dimensions and twenty subcategories, each with explicit, detailed criteria for every score from 1 to 5 (Table~\ref{tab:rar_rubric}). This structure minimizes ambiguity and subjectivity, supporting transparent and reproducible evaluation for both human- and LLM-generated surveys. All criteria and scoring anchors are fully documented (Appendix~2), providing a robust foundation for benchmarking, actionable feedback, and community adaptation to new domains.

\begin{table}[t]
\centering
\small
\setlength{\tabcolsep}{0.5mm}
\begin{tabular}{ll}
\toprule
\textbf{Dimension} & \textbf{Subcategories} \\
\midrule
Scope         & Objectives, Relevance, Audience \\
Literature    & Comprehensiveness, Balance, Currency \\
Analysis      & Depth, Integration, Gaps \\
Originality   & Novelty, Advancement, Redundancy Avoidance \\
Organization  & Logical Flow, Section Clarity, Summarization \\
Presentation  & Language, Visuals, Formatting \\
References    & Accuracy, Appropriateness \\
\bottomrule
\end{tabular}
\caption{Structure of the shared evaluation rubric: seven dimensions, twenty subcategories, each scored from 1 to 5 (total 100 points).}
\label{tab:rar_rubric}
\end{table}

\section{Experiment Setup}
\label{sec:experiments}

\subsection{Benchmarking and Comparison Framework}
\subsubsection{Human-Written Baselines} 
We benchmark ARISE against ten recently published, human-authored survey papers, selected for topical diversity, publication recency (2023–2025), and high visibility in either arXiv or leading peer-reviewed venues. These baselines were chosen to represent a spectrum of research areas—including LLM reasoning, evaluation, multimodality, time-series modeling, and human-agent interaction—ensuring relevance to our target domains. See Appendix 3 for the full summary table of baseline survey papers. For direct comparison, we use ARISE to generate survey papers on the same research areas as the human-written baselines.

\subsubsection{Automated Survey Generation Systems} 
We also evaluate ARISE against prior automated systems, including \textit{SurveyForge}~\cite{ref12}, \textit{SurveyX}~\cite{ref11}, and \textit{AutoSurvey}~\cite{ref10}. Due to availability constraints, we include 10 papers each from SurveyForge and SurveyX, and 3 from AutoSurvey.

\subsection{Experimental Design }
ARISE is implemented using CrewAI~\cite{crewai2024}, with API keys for GPT-4.1~\cite{openai2023gpt4}, Gemini 2.5 Pro~\cite{geminiteam2025geminifamilyhighlycapable}, and Claude 3.7 Sonnet~\cite{anthropic2024claude} managed via environment variables. All system agents run on GPT-4.1 by default. Our pipeline also integrates the Serper API~\cite{serper2024} for web search and source scraping(see Appendix 5 for cost and time details). The rubric, shown in Table~\ref{tab:rar_rubric}, defines twenty subcategories across seven core dimensions, and all reported scores are computed using the aggregate formula given in Eq.~\eqref{eq:average_score}. 

To ensure full-document coverage and minimize positional bias, we segment each paper into contiguous 3-page chunks, encouraging balanced attention across sections and ensuring each chunk fits within the LLMs’ context windows. Chunk-level reviews produce localized rubric scores, enabling fine-grained analysis of writing quality and document structure. All outputs from ARISE, competing automated systems, and human-written baselines are evaluated using the same domain-agnostic rubric, chunking strategy, and tri-model reviewer setup.

Beyond automated evaluation, we include a human expert study to validate perceived quality improvements after iterative refinement. We also perform a model-based ablation, comparing end-to-end system performance with both small (\texttt{gpt-4.1-mini}) and large (\texttt{gpt-4.1}) language models as pipeline agents. Finally, we conduct a citation traceability audit to assess reference reliability and report inter-rater agreement using Krippendorff’s Alpha.

\section{Results and Analysis}
\label{sec:results}
\subsection{Overall Performance.}
Our system, \textbf{ARISE}, achieves the highest overall quality among all five evaluated systems. As shown in Table~\ref{tab:mean_total_scores}, On our agentic rubric-based evaluation framework, ARISE’s outputs receive higher scores than all baselines across each individual reviewer—Claude 3.7, Gemini 2.5 Pro, and GPT-4.1—with a tri-judge average of \textbf{92.48} (Table~\ref{tab:mean_total_scores}). To check for potential self-judging bias, we also report a bi-judge setting that \emph{excludes} GPT-4.1 from the reviewer pool (Bi-judge (G+C) column); ARISE remains the top-performing system (92.43 vs.\ the next best 87.58), indicating that our conclusions do not depend on using the generator family as a judge.

\begin{table}[t]
\centering
\small
\setlength{\tabcolsep}{1.2pt} 
\begin{adjustbox}{max width=\linewidth}
\begin{tabular}{lccccc}
\toprule
System & GPT-4.1 & Gemini & Claude & Tri-judge Avg & Bi-judge Avg  \\
\midrule
\textbf{ARISE}      & \textbf{92.57} & \textbf{92.59} & \textbf{92.27} & \textbf{92.48} & \textbf{92.43} \\
AutoSurvey          & 81.87          & 81.74          & 83.76          & 82.46          & 82.75          \\
Baseline     & 86.58          & 85.81          & 86.06          & 86.15          & 85.94          \\
SurveyForge         & 87.88          & 87.51          & 87.64          & 87.68          & 87.58          \\
SurveyX             & 81.13          & 81.99          & 81.82          & 81.65          & 81.91          \\
\bottomrule
\end{tabular}
\end{adjustbox}
\caption{
Mean TOTAL scores by system and reviewer.
“Tri-judge Avg” uses all three reviewers (GPT-4.1, Gemini 2.5 Pro, Claude 3.7 Sonnet).
“Bi-judge Avg (G+C)” excludes the generator family and averages only Gemini 2.5 Pro and Claude 3.7 Sonnet.
\label{tab:mean_total_scores}
}
\end{table}

\subsection{Rubric-Level Superiority}

ARISE demonstrates consistently strong rubric-level performance, leading in all seven categories evaluated (Table~\ref{tab:MeanReviewers_by_category}). In critical dimensions such as \textit{Literature} (4.95), \textit{Presentation} (4.84), \textit{References} (4.98), and \textit{Organization} (4.82), ARISE surpasses both human-written baselines and recent automated systems. These gains are attributed to our agentic refinement strategy and explicit modular writing design.

\begin{table}[!htbp]
\centering
\setlength{\tabcolsep}{0.5mm}
\small
\begin{tabular}{l@{\hspace{0.5mm}}ccccc}
\toprule
Category   & \textbf{ARISE} & Autosurvey & Baseline & SurveyForge & SurveyX \\
\midrule
Analysis      & \textbf{4.56}  & 3.94      & 3.74     & 4.45        & 3.64 \\
Literature    & \textbf{4.95}  & 4.53      & 4.70     & 4.79        & 4.29 \\
Organization  & \textbf{4.82}  & 4.26      & 4.52     & 4.55        & 4.41 \\
Originality   & \textbf{4.44}  & 3.88      & 3.97     & 4.10        & 3.68 \\
Presentation  & \textbf{4.84}  & 3.57      & 4.35     & 3.87        & 4.22 \\
References    & \textbf{4.98}  & 4.81      & 4.94     & 4.85        & 4.45 \\
Scope         & \textbf{4.30}  & 4.10      & 4.14     & 4.23        & 4.00 \\
\bottomrule
\end{tabular}
\caption{Mean reviewer scores by rubric category and system}
\label{tab:MeanReviewers_by_category}
\end{table}

ARISE’s consistently high scores across all rubric dimensions confirm the effectiveness of our modular agent design and rubric-guided refinement strategy.  Inter-rater reliability among the reviewers was consistently high across all systems, with Krippendorff’s Alpha (\(\alpha\)) exceeding \textbf{0.966} in all cases and reaching up to \textbf{0.987}—see Appendix 4 for full agreement scores.

These results validate our core hypothesis: rubric-guided iterative refinement enables transparent, interpretable, and high-quality survey generation. By integrating modular agent roles, structured evaluation criteria, and multi-round refinement, ARISE achieves higher rubric scores than baseline and other generation systems, and establishes a reproducible foundation for self-improving academic writing.

\subsection{Validating the Refinement Process with a Domain-Specific Example}

To illustrate the impact of ARISE’s rubric-guided iterative refinement loop, we present a representative refinement trajectory for a system-generated survey paper in the domain of \textit{LLM reasoning and replication}. This example tracks how quality evolves over successive refinement iterations based on rubric-guided feedback from three independent reviewer agents.

\begin{table}[t]
\centering
\setlength{\tabcolsep}{0.5mm} 
\small
\begin{tabular}{lcccc}
\toprule
\textbf{Round} & \textbf{GPT-4.1} & \textbf{Gemini 2.5 Pro} & \textbf{Claude 3.7 Sonnet} & \textbf{Average} \\
\midrule
0 & 86.8 & 87.9 & 86.2 & 87.0 \\
1 & 90.5 & 89.8 & 90.0 & 90.1 \\
2 & 93.8 & 89.2 & 91.3 & 91.4 \\
3 & 92.3 & 92.8 & 92.9 & 92.7 \\
\bottomrule
\end{tabular}
\caption{A case of average review score progression across rubric-guided refinement rounds for a generated survey. The target average score is 92.0.}
\label{tab:self_improvement_rounds_avg}
\end{table}

As shown in Table~\ref{tab:self_improvement_rounds_avg}, the case begins with an average reviewer score of 87.0. By Round 3, the average score exceeds the threshold, reaching 92.7. This trajectory demonstrates ARISE’s ability to iteratively elevate content quality through modular, reviewer-guided revision. See the supplementary material for results from all other topic domains.

\subsection{Human Evaluation}

To assess the effectiveness of our agentic refinement process, we conducted a human evaluation with four experts (two professors, one postdoc, one PhD student). Each expert independently reviewed five system-generated survey papers, both \emph{before} and \emph{after}  iterative refinement, using the same 20-subcategory rubric as in our automated evaluation.

\paragraph{Results.} Table~\ref{tab:human_eval} summarizes the scores. On average, the total score increased from 70.2 to 83.7 out of 100 , and the mean subcategory rating rose from 3.51 to 4.18 out of 5. All topics and reviewers observed substantial, consistent improvement. After refinement, system outputs consistently achieved ``strong'' ($\geq$4.0) expert ratings.

\begin{table}[ht]
\centering
\small
\setlength{\tabcolsep}{1 mm}
\begin{tabular}{lcccc}
\toprule
\textbf{Topic} & \textbf{Before} & \textbf{After} & \textbf{Before} & \textbf{After} \\
 & (Total) & (Total) & (Avg) & (Avg) \\
\midrule
ASG LitRev & 69.8 & 80.5 & 3.49 & 4.03 \\
GenAI in Manufacturing & 68.8 & 82.8 & 3.44 & 4.14 \\
LLM Reasoning & 66.3 & 86.0 & 3.31 & 4.30 \\
Clustering Indexing & 70.5 & 82.0 & 3.53 & 4.10 \\
Retrieval-Aug. Gen. & 75.8 & 87.0 & 3.79 & 4.35 \\
\midrule
\textbf{Average} & 70.2 & 83.7 & 3.51 & 4.18 \\
\textbf{Improvement (\%)} & -- & \textbf{19.2\%} & -- & \textbf{19.2\%} \\

\bottomrule
\end{tabular}
\caption{Human evaluation scores for five system-generated surveys (N=4 experts per paper). ``Total'' is out of 100; ``Avg'' is per-rubric average out of 5.}
\label{tab:human_eval}
\end{table}

These results confirm that rubric-guided refinement substantially and reliably improves survey quality as perceived by domain experts. Additional human evaluations are detailed extensively in Appendix 6.

\subsection{Model-Based Ablation Study}

To assess how agent capacity influences end-to-end survey generation quality, we conducted an ablation study using two full system configurations: one with \texttt{gpt-4.1-mini} as the primary agent in all modules, and another with the larger \texttt{gpt-4.1}. For each configuration, we generated complete survey papers and evaluated them across refinement rounds using the same rubric-guided iterative process.

As shown in Table~\ref{tab:model_eval_avg}, both configurations demonstrate quality improvements through the iterative refinement loop. The system powered by the larger model achieved a higher final rubric score and greater improvement, reflecting the advantages of increased model capacity for both generation and refinement tasks. Even with the smaller model configuration, ARISE achieves an average final rubric score of \textbf{88.04}, which remains slightly higher than those of prior automated systems and human-written baselines, as shown in Table~\ref{tab:mean_total_scores}.

\begin{table}[ht]
\centering
\small
\begin{tabular}{lcccc}
\toprule
\textbf{System Model} & \textbf{Initial} & \textbf{Final} & \textbf{Improvement (\%)} \\
\midrule
\texttt{gpt-4.1-mini} & 83.09 & 88.04 & \textbf{5.96\%} \\
\texttt{gpt-4.1}   & 86.53 & 92.48  & \textbf{6.88\% }\\
\bottomrule
\end{tabular}
\caption{Average per-reviewer rubric score across refinement rounds for system-generated papers using small and large models.}
\label{tab:model_eval_avg}
\end{table}

\subsection{ Reference Reliability}

To validate the reliability of ARISE’s citation preparation process, we conducted a traceability audit of the final system-generated references. We define the \textit{Expanded Citation Traceability Rate} (eCTR) as:

\begin{equation}
\text{eCTR} = \frac{V}{T}, \quad
\text{Hallucination Rate} = 1 - \text{eCTR}
\label{eq:ectr}
\end{equation}
\noindent
where $V$ is the number of verifiable citations successfully matched to external databases, and $T$ is the total number of citations extracted from the system-generated PDF.

We applied layout-aware reference extraction (via PyMuPDF) to final PDF outputs and matched each citation against CrossRef, Semantic Scholar, and arXiv using public APIs. Across all evaluated ARISE outputs, we observed a perfect mean eCTR of \textbf{1.00}, corresponding to a hallucination rate of \textbf{0.00}. This result demonstrates the robustness of our citation-first pipeline in producing factually grounded scholarly references.

\section{Conclusion}
\label{sec:conclusion}

We present \textsc{ARISE}, a modular, agentic system for automated academic survey generation and iterative refinement. By decomposing the survey writing and peer review process into specialized LLM-powered agents, ARISE delivers transparent, reproducible, and high-quality scholarly outputs, and effectively addressing longstanding challenges in quality control, formatting, and iterative improvement. Our experiments show that ARISE achieves higher rubric scores than both human-written baselines and state-of-the-art automated survey generation systems across a comprehensive, behaviorally anchored evaluation rubric. The system demonstrates near-perfect reference reliability and substantial quality improvements through rubric-guided, multi-agent iterative refinement. Further discussion of broader impact, ethical statement, and limitations is provided in Appendix 5.

\appendix
\renewcommand\thetable{A\arabic{table}}
\renewcommand\thefigure{A\arabic{figure}}
\setcounter{table}{0}
\setcounter{figure}{0}

\section*{Appendix}

\section*{1. Rubric-Guided Iterative Refinement Sample}
\label{app:arl-feedback}

Each round of refinement in our framework is guided by structured feedback from multiple reviewer agents, who independently evaluate each manuscript draft using a detailed, standardized rubric (see Table~\ref{tab:rubric_detailed}). Agents provide both quantitative scores and qualitative suggestions; representative reviewer outputs and sample revision recommendations are compiled in Tables~\ref{tab:arl-feedback}--\ref{tab:review_summary_grouped}. These records illustrate the diversity and depth of agentic assessment throughout the iterative improvement process described in the main text. 

For full transparency and reproducibility, all generated survey drafts, reviewer feedback, meta-review synthesis, and the codebase for executing rubric-driven evaluation and revision are provided in the supplementary materials (see each topic's subfolder \texttt{review\_output}).

\begin{table*}[t]
\centering
\begin{tabularx}{\textwidth}{p{2.5cm} p{4cm} X}
\toprule
\textbf{Model} & \textbf{Section} & \textbf{Suggestions} \\
\midrule
gpt-4.1 & Abstract, Introduction, and Historical and Foundational Landscape & Expand literature coverage, especially in the benchmarking and reasoning evaluation literature.; Ensure all referenced tables and visuals are present, clear, and visually improve synthesis.; Replace generic numbered citations with a full reference list in the final version for traceability. \\
\addlinespace
gemini-2.5 & Abstract, Introduction, Historical and Foundational Landscape (through start of Benchmarking) & Ensure referenced figures/tables (e.g., Table 1) are included in the final version.; Verify full references section for accuracy and formatting.; Strengthen in-chunk summarization using inline tables or boxes if possible to reinforce key comparative points. \\
\addlinespace
claude-3.7 & Introduction, Historical and Foundational Landscape & Broaden and deepen the engagement with competing or alternative views where appropriate (e.g., critiques of hybrid models or transformer approaches).; Ensure that figures, tables, and diagrams are present and directly support claims when referenced.; Replace placeholder citation markers with complete bibliography for full submission.; Consider summarizing key takeaways at the end of major sections more explicitly. \\
\addlinespace
gpt-4.1 & 3.2–4.3 Benchmark Evaluation and Probing Sections & Add an explicit restatement or recap of the overall survey objectives when introducing new major sub-sections.; Provide a few concrete examples where benchmarking volatility misled the field (to deepen critical analysis).; Highlight implications or actionable guidance for benchmark and metric developers.; Consider briefly summarizing emerging benchmarks from late 2023 or 2024, if possible, for currency. \\
\addlinespace
gemini-2.5 & 3. Benchmarking, Evaluation, and Comparative Analysis & Provide a brief, explicit statement of objectives at the start or end of the section.; Consider enhancing section summaries or explicitly restating takeaways after major analyses.; Integrate more conceptual figures to complement the empirical tables.; Check for correct and non-redundant formatting in citation numbering. \\
\addlinespace
claude-3.7 & Benchmarking and Evaluation Paradigms; Probing, Reasoning, and Linguistic Benchmarks & Add an explicit section-level objective statement or overview at the start.; Improve section transitions or provide mini-introductions to major subsections.; Standardize citation formatting in text and ensure consistent reference styling.; Consider including workflow diagrams, paradigm maps, or conceptual illustrations.; Make audience/who-will-benefit aspects clear in introductory text. \\
\bottomrule
\end{tabularx}
\caption{Rubric-Guided Iterative Refinement Sample}
\label{tab:arl-feedback}
\end{table*}

\renewcommand{\arraystretch}{1.15}

\begin{table*}[t]
\centering
\small
\begin{adjustbox}{width=\textwidth}
\begin{tabularx}{\textwidth}{l l l l l l X}
\toprule
\textbf{Model} & \textbf{Chunk} & \textbf{ChunkIndex} & \textbf{Category} & \textbf{Sub-Criterion} & \textbf{Score} & \textbf{Comment (shared for category)} \\
\midrule

gpt-4.1 & Pages 1--3 & 0 & Literature & Comprehensiveness & 4 & The section references many key works across subfields and uses over 30 citations (references numbering up to 114 are mentioned), but the specific breadth cannot be fully verified as most citation details are numeric placeholders in this excerpt. Coverage appears current, with references to recent methods and issues, but explicit years or full coverage cannot be absolutely confirmed. Discussion is balanced and cites limitations and open problems. \\
gpt-4.1 & Pages 1--3 & 0 & Literature & Balance & 5 &  \\
gpt-4.1 & Pages 1--3 & 0 & Literature & Currency & 4 &  \\
\midrule

gemini-2.5 & Pages 1--3 & 0 & Literature & Comprehensiveness & 5 & The section displays strong comprehensiveness, referencing foundational and very recent works (with $\geq$ 30 citations), covering multiple domains (survey generation, LLMs, hybrid workflows). Balance is maintained by discussing benefits and challenges, though the critique could be more granular. Currency is excellent, with up-to-date references. \\
gemini-2.5 & Pages 1--3 & 0 & Literature & Balance & 4 & \\
gemini-2.5 & Pages 1--3 & 0 & Literature & Currency & 5 & \\
\midrule

claude-3.7 & Pages 1--3 & 0 & Literature & Comprehensiveness & 5 & References are numerous and span many subfields, often citing multiple recent papers in each subtopic ($\geq$30 citations appear, many recent, referencing preprints/conferences). Strengths and weaknesses are discussed, including presenting trade-offs (e.g., hybrid vs. black-box). Multiple perspectives are mentioned, but direct debates between competing schools are limited. \\
claude-3.7 & Pages 1--3 & 0 & Literature & Balance & 4 & \\
claude-3.7 & Pages 1--3 & 0 & Literature & Currency & 5 & \\
\bottomrule
\end{tabularx}
\end{adjustbox}
\caption{Detailed rubric scores sample of Literature category for reviewer agents (Pages 1--3, Chunk 0), grouped by agent with shared comments.}
\label{tab:appendix_rubric_detail}
\end{table*}

\renewcommand{\arraystretch}{1.13}
\setlength{\tabcolsep}{3pt}

\begin{table*}[t]
\centering
\small
\begin{adjustbox}{width=\textwidth}
\begin{tabularx}{\textwidth}{
  l l
  >{\raggedright\arraybackslash}p{3.5cm}   
  >{\raggedright\arraybackslash}p{3.2cm}   
  >{\raggedright\arraybackslash}p{2.6cm}   
  >{\raggedright\arraybackslash}p{2.7cm}   
  >{\centering\arraybackslash}p{0.6cm}     
  >{\centering\arraybackslash}p{1.1cm}     
}
\toprule
\textbf{Model} & \textbf{Chunk} & \textbf{Summary} & \textbf{Major Strengths} & \textbf{Major Weaknesses} & \textbf{Recommendations} & \textbf{Final Score} & \textbf{Dec.} \\
\midrule

gpt-4.1 & Pages 1--3
& This section is exemplary for an academic survey, setting a strong foundation for both technical insight and methodological rigor. It weaves together the latest advances in agentic systems, AI-driven survey automation, and academic workflow innovation, making substantial contributions via new taxonomies, thorough discussion of standards, and explicit identification of unresolved challenges.
& Explicit objectives and robust relevance to urgent AI and academia intersections; Integrated, up-to-date, and balanced literature coverage; Thorough analysis, new frameworks, and clearly presented challenges/gaps; Excellent organization and professional presentation.
& Minor gaps in visual diversity—additional figures would be valuable; Clustered citations could be better detailed in the references section.
& Add one or more schematic diagrams or flowcharts of workflow architectures; Maintain detailed, accessible reference lists for full citation clarity.
& 4.9
& accept
\\
\addlinespace

gemini-2.5 & Pages 1--3
& This section of the survey provides a comprehensive, state-of-the-art review of AI-driven academic automation. Theoretical and workflow foundations are robust, and major research challenges are well-delineated via succinct tables. The taxonomy of GAEs and focus on explainability, standardization, and equity are notable advances.
& Clear objectives and targeted audience; Thorough, recent, and balanced literature review; Insightful theoretical foundation and comparative analysis; Succinct, effective summarization via tables
& Minor overlap with prior automation/LLM surveys; Occasional imbalance in strength vs weakness discussion; Visuals could extend beyond tables; citation formatting is numeric-only
& Minor revision: emphasize weaknesses/limitations alongside strengths more uniformly; Introduce more diverse visuals (e.g., workflow diagrams); Include reference list or mapping in the supplementary material for citation traceability
& 4.7
& minor revision
\\
\addlinespace

claude-3.7 & Pages 1--3
& This is an ambitious and well-written survey that provides a timely, comprehensive, and insightful overview of automated survey generation, literature review, and agentic academic AI systems. The writing is clear and professional, objectives are explicit, and the structure is logical and effective. Integration of tables reinforces important points and summarizes complex ideas. The taxonomy of GAEs is novel and enhances the field. Citations and literature engagement are mature but could be marginally more explicit or current in places.
& Explicit objectives and well-defined scope; Addresses cutting-edge trends and pressing academic needs; Original taxonomy/framework and clear advancement for the field; Balanced, analyzed, and critical discussion; Effective summarization and organization
& Some integration could be strengthened between technical and methodological examples; Citation formatting and clarity could be made more consistent; Minor need for additional visuals to complement tables
& Strengthen interdisciplinary connections and integration across subfields; Standardize and clarify citation format throughout the section; Incorporate additional figures or diagrams for workflow illustration; Elaborate on specific research gaps with suggested future directions
& 4.6
& minor revision
\\
\bottomrule
\end{tabularx}
\end{adjustbox}
\caption{Grouped review summary scores for each section (Pages 1--3 and 4--6) and agent.}
\label{tab:review_summary_grouped}
\end{table*}

\subsection{Generated Survey Sample}

\begin{figure*}[htbp]
    \centering
    \includegraphics[width=\textwidth]{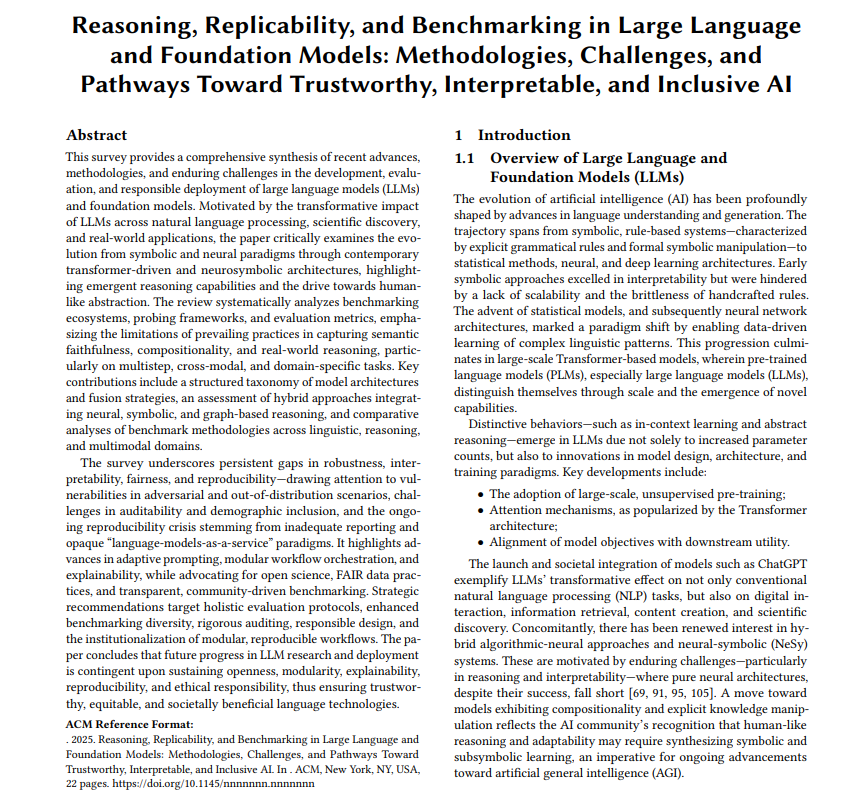}
    \caption{
      Example first page of a survey paper generated by ARISE, formatted in ACM style. This sample illustrates the structured abstract, clear academic formatting, and section organization achieved by the pipeline. Full outputs for additional topics are provided in the supplementary materials.
    }
    \label{fig:sample_paper}
\end{figure*}

Our ARISE system assists academic researchers by generating high-quality PDF drafts using the full \texttt{TeX Live} library, which supports a wide variety of \LaTeX\ classes for major journals and conferences (e.g., AAAI, ACM, IEEE, Springer, Elsevier, and more). The title and abstract for each draft are autonomously generated by dedicated agents based on the complete manuscript content, ensuring that each paper features a contextually relevant and well-structured summary. The modular formatting and finalization module enables users to specify their preferred style, automatically adapting the manuscript structure, citation format, and page layout to the conventions of the target venue. ARISE manages all technical aspects of \LaTeX\ formatting, including title, abstract, metadata, tables, and references, allowing researchers to focus on the intellectual content rather than formatting details. This workflow streamlines the preparation of survey drafts in \LaTeX, saving substantial time and effort when initiating new research projects. An example output generated by ARISE is illustrated in Figure~\ref{fig:sample_paper}.

\subsection{Reference Sample}
\begin{figure*}[htbp]
    \centering
    \includegraphics[width=\textwidth]{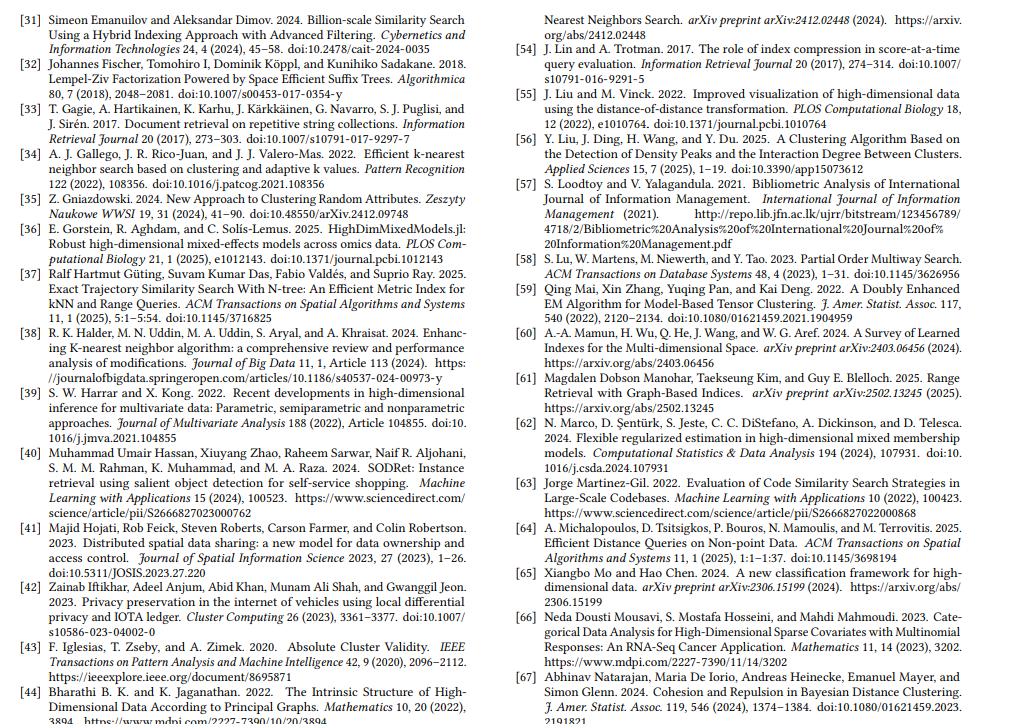}
    \caption { 
    Reference sample generated by ARISE on the topic of \textit{Clustering, Indexing, and Data Structures for High-Dimensional and Categorical Data}, highlighting its prioritization of peer-reviewed journal and conference papers to ensure citation quality and credibility.}
    \label{fig:sample_references}
\end{figure*}

To illustrate the professional formatting quality and citation currency of ARISE-generated manuscripts, we provide a reference sample generated by our system on the topic of \textit{Clustering, Indexing, and Data Structures for High-Dimensional and Categorical Data} (see Figure~\ref{fig:sample_references}).

ARISE emphasizes citations from peer-reviewed journals and established venues when available, while still incorporating relevant preprints for timeliness. This approach enhances the credibility, accuracy, and academic rigor of the generated surveys compared to systems predominantly reliant on preprints.

\section*{2. Rubric Structure and Scoring Anchors.}
To operationalize consistent, high-fidelity review, Table~\ref{tab:rubric_detailed} presents the complete scoring rubric $\mathcal{B}$ used throughout our agentic evaluation pipeline. Each dimension and subcategory is defined with explicit, behaviorally anchored criteria for every possible score from 1 (lowest) to 5 (highest). This granular structure—covering scope, literature review, analysis, originality, organization, presentation, and references—translates general peer-review standards into actionable, reproducible evaluation guidance for both human and automated assessment. By making all criteria and scoring anchors fully explicit, the rubric supports transparent benchmarking, facilitates community adaptation, and underpins robust feedback cycles in our iterative refinement framework.

\begin{table*}[t]
\centering
\scriptsize
\begin{adjustbox}{width=\textwidth}
\begin{tabular}{
  >{\raggedright\arraybackslash}p{1.2cm}
  >{\raggedright\arraybackslash}p{1.8cm}
  >{\raggedright\arraybackslash}p{2.0cm}
  >{\raggedright\arraybackslash}p{2.3cm}
  >{\raggedright\arraybackslash}p{2.2cm}
  >{\raggedright\arraybackslash}p{2.2cm}
  >{\raggedright\arraybackslash}p{2.2cm}
}
\toprule
\textbf{Category} & \textbf{Criterion} & \textbf{1} & \textbf{2} & \textbf{ 3} & \textbf{ 4} & \textbf{ 5} \\
\midrule
Scope & Objectives & No objectives stated or inferred & Unclear or implicit; requires inference & Vague or generic; lacks focus & Clear in one section; lacks precision & Clearly stated in abstract and intro; scoped and measurable \\
& Relevance & Not relevant to the field & Weak or outdated connection & Partially related to broader topic & Generally relevant, not urgent & Directly aligns with high-impact trends \\
& Audience & No discernible audience & Confusing or poorly targeted & Somewhat unclear & Generally appropriate tone & Clear academic or interdisciplinary targeting \\
\midrule
Literature & Comprehensiveness & Sparse or incomplete coverage & Major omissions & Some omissions or limited domain & Mostly complete with minor gaps & $\geq$ 30 citations, across subfields, up-to-date \\

& Balance & Highly biased or promotional & One-sided view & Somewhat unbalanced & Balanced with minor bias & Discusses strengths/weaknesses and perspectives \\
& Currency & Ignores recent developments & Mostly dated content & Some outdated dominance & Mostly recent with few older works & Up-to-date including preprints and conferences \\
\midrule
Analysis & Depth & No meaningful analysis & Minimal or weak analysis & Descriptive only & Moderate depth & Theoretical rigor, layered insight \\
& Integration & Disjointed and fragmented & Mostly disconnected ideas & Partial, siloed integration & Good integration & Seamless integration of multiple perspectives \\
& Gaps & Ignores all research gaps & Barely addresses open questions & Surface-level mention & Mentions some gaps & Clearly identifies open challenges \\
\midrule
Originality & Novelty & No original contribution & Mostly derivative & Slightly original & Novel combination of ideas & New taxonomy, framework, or domain \\
& Advancement & No advancement & Minimal progress & Incremental value & Moderate contribution & Strong guidance for future research \\
& Redundancy Avoidance & Highly repetitive & Largely redundant & Moderate overlap & Mostly unique & Clearly distinct from prior surveys \\
\midrule
Organization & Logical Flow & Chaotic and disorganized & Poor transitions & Basic structure with issues & Mostly clear flow & Excellent transitions and structure \\
& Section Clarity & No clear structure & Unclear or unlabeled & Confusing or too long & Mostly clear & Well-labeled and crystal clear \\
& Summarization & No summary or synthesis & Almost none & Minimal synthesis & Some synthesis and structure & Effective use of summaries and visuals \\
\midrule
Presentation & Language & Unreadable or ungrammatical & Poor grammar or clarity & Clumsy tone & Mostly well-written & Clear academic language throughout \\
& Visuals & No meaningful visuals & Irrelevant or low-quality & Basic, not integrated & Good visuals with minor issues & Strong figures/tables supporting content \\
& Formatting & Disorganized formatting & Distracting issues & Inconsistent formatting & Minor format problems & Clean, consistent styles \\
\midrule
References & Accuracy & Unreliable or incorrect citations & Multiple citation errors & Some mismatched or incomplete & Minor format issues & Accurate, traceable, properly formatted \\
& Appropriateness & Poor citation quality & Many low-quality sources & Some irrelevant or filler & Mostly appropriate & Highly relevant, current and foundational \\
\bottomrule
\end{tabular}
\end{adjustbox}
\caption{Evaluation Rubric for Survey Paper Quality (Scores 1--5)}
\label{tab:rubric_detailed}
\end{table*}

\section{3. Baseline Survey Papers Used for Benchmarking}
\label{app:baseline_topics}

To ensure fair and meaningful benchmarking, the selection of baseline research domains was constrained by the availability of comparable sample outputs from all three automated survey generation systems evaluated in this study. Within these feasible domains, we aimed for a balanced representation by including five preprint surveys and five peer-reviewed surveys. Table~\ref{app:baseline_topics} details each baseline’s research area, full title with citation, and publication venue or year, providing a transparent reference set for direct comparison against ARISE and prior automated systems.

\begin{table*}[!htbp]
\centering
\label{tab:baseline_topics}
\begin{adjustbox}{width=0.98\linewidth}
\begin{tabular}{p{0.32\textwidth} p{0.40\textwidth} p{0.22\textwidth}}
\toprule
\textbf{Research Area} & \textbf{Paper Title and Reference} & \textbf{Venue / Year} \\
\midrule
LLM reasoning and replication & \textit{100 Days After DeepSeek-R1: A Survey on Replication Studies...}~\cite{zhang2025100daysdeepseekr1survey} & arXiv 2025 \\
LLM evaluation metrics & \textit{A Survey on Evaluation of Large Language Models}~\cite{chang2023surveyevaluationlargelanguage} & ACM 2024 \\
Retrieval-augmented generation & \textit{Retrieval-Augmented Generation for LLMs: A Survey}~\cite{gao2024retrievalaugmentedgenerationlargelanguage} & arXiv 2024 \\
Multimodal large language models & \textit{A Survey on Multimodal Large Language Models}~\cite{Yin_2024} & NSR 2024 \\
Time-series modeling & \textit{Time-Series Large Language Models: A Systematic Review of State-of-the-Ar}~\cite{DBLP:journals/access/AbdullahiDZAZA25} & IEEE 2025 \\
Generative AI in manufacturing & \textit{Generative Machine Learning in Adaptive Control of Dynamic Manufacturing Processes: A Review}~\cite{lee2025generativemachinelearningadaptive} & arXiv 2025 \\
Topology of fractal squares & \textit{A Survey on the Topology of Fractal Squares}~\cite{luo2025surveytopologyfractalsquares} & arXiv 2025 \\
Human-agent interaction & \textit{Humanizing llms: A survey of psychological measurements with tools, datasets, and human-agent applications}~\cite{dong2025humanizingllmssurveypsychological} & arXiv 2025 \\
Disease detection across modalities & \textit{A Methodological and Structural Review of Parkinson’s Disease Detection Across Diverse Data Modalitie}~\cite{miah2025methodologicalstructuralreviewparkinsons} & IEEE 2025 \\
Generative AI in mobile networks & \textit{Generative AI in Mobile Networks: A Survey}~\cite{karapantelakis2024generative} & AoT 2024 \\
\bottomrule
\end{tabular}
\end{adjustbox}
\caption{Selected Baseline Survey Papers with Research Areas and Publication Venues}
\label{app:baseline_topics}
\end{table*}

\section{4. Reliability Evaluation}
\label{app:reliability}

To assess consistency among reviewer agents, we compute \textbf{Krippendorff’s Alpha} (\(\alpha\)), a standard inter-rater reliability metric. It is defined as:
\[
\alpha = 1 - \frac{D_o}{D_e}
\]
where \(D_o\) denotes observed disagreement and \(D_e\) denotes expected disagreement by chance. Values range from \(-\infty\) to 1, with \(\alpha = 1\) indicating perfect agreement.

We compute \(\alpha\) under the interval-level setting using rubric scores from three reviewers (GPT-4.1, Gemini 2.5, Claude 3.7) across four systems and a published baseline. All evaluations use the same rubric and chunking protocol.

\begin{table}[h]
\centering
\begin{adjustbox}{width=\linewidth}
\begin{tabular}{llc}
\toprule
\textbf{Type} & \textbf{System / Model Pair} & \textbf{Krippendorff’s Alpha (\(\alpha\))} \\
\midrule
\multicolumn{3}{l}{\textit{System-Level Agreement}} \\
& SurveyX                & 0.974 \\
& SurveyForge            & 0.977 \\
& Baseline (Published)   & 0.966 \\
& Autosurvey             & 0.987 \\
& ARISE                  & 0.973 \\
\midrule
\multicolumn{3}{l}{\textit{Model-Level Agreement (All Systems)}} \\
& Claude 3.7 vs Gemini 2.5 & 0.974 \\
& Claude 3.7 vs GPT-4.1    & 0.977 \\
& Gemini 2.5 vs GPT-4.1    & 0.973 \\
\bottomrule
\end{tabular}
\end{adjustbox}
\caption{Krippendorff’s Alpha scores for system-level and inter-model agreement. Computed using interval-scale rubric ratings across reviewer agents.}
\label{tab:krippendorff-combined}
\end{table}

As shown in Table~\ref{tab:krippendorff-combined}, all systems—including ARISE, the baselines, and prior automated methods—achieved exceptionally high inter-rater reliability, with Krippendorff’s Alpha (\(\alpha\)) values exceeding 0.96 across the board. This reflects strong consistency in scoring among the independent reviewer agents and further supports the robustness and interpretability of the rubric-guided evaluation protocol adopted in our study. High agreement at both the system and model levels suggests that our rubric design and agent workflow produce reproducible, reliable assessments suitable for benchmarking survey generation quality.

\section{5. Limitations, Broader Impact, and Ethical Statement}
\label{sec:Limitations}
\subsection*{Limitations}
While ARISE achieves strong empirical performance in generating and evaluating survey papers, several limitations should be acknowledged.

First, our evaluation framework relies entirely on large language models (LLMs) as reviewer agents—specifically GPT-4.1, Gemini 2.5 Pro, and Claude 3.7 Sonnet. Although we adopt a detailed and standardized rubric to promote consistency, we were unable to involve human reviewers due to time and resource constraints. As such, the evaluation may reflect alignment patterns and blind spots specific to current LLMs.

Second, Our system relies on commercial LLM APIs with tiered pricing and computational constraints, such as request rate limits, context window caps, and quota exhaustion. These factors can intermittently affect pipeline stability or trigger retry logic. We employ both large-scale models (e.g., GPT-4.1, Claude 3.7, Gemini 2.0 Pro) and smaller, cost-efficient alternatives (e.g., GPT-4.1 Mini, Gemini Flash), balancing performance and budget. As shown in Figure~\ref{tab:llm-pricing}, larger models tend to offer stronger reasoning and editing capabilities but at significantly higher costs—up to 5–10× per million tokens compared to compact variants. In addition to LLM usage, ARISE integrates paid external APIs such as Serper for web search and scraping, which incurs \$0.01 per query beyond the free tier (100/month), making it affordable at an academic scale. For our experiments, total Serper usage cost remained under \$200.

In practice, a single paper generation cycle typically costs \$10–\$20 and takes approximately 3.5 hours, depending on the topics. As topic complexity and breadth grow, additional time is required for citation expansion, data collection, and outline refinement. Notably, our refinement module—which includes agentic evaluation and iterative rewriting—accounts for 30-40 percent of total processing time, especially when multi-round improvement is needed for quality assurance.

\begin{table}[htbp]
\centering
\begin{tabular}{llccc}
\toprule
\textbf{Model} & \textbf{Type} & \textbf{Input} & \textbf{Output} & \textbf{Context Caching} \\
\midrule
GPT-4.1 & Premium & \$2.00 & \$8.00 & \$0.50 \\
GPT-4.1 mini & Balanced & \$0.40 & \$1.60 & \$0.10 \\
Claude 3.7 Sonnet & Premium & \$3.00 & \$15.00 & \$0.30 (read), \$3.75 (write) \\
Gemini 2.5 Pro & Premium & \$1.25 / \$2.50 & \$10.00 / \$15.00 & \$0.31 / \$0.625 \\
Gemini 2.0 Flash & Affordable & \$0.10 (text/image), \$0.70 (audio) & \$0.40 & \$0.025 \\
\bottomrule
\end{tabular}
\caption{LLM Pricing Comparison (USD per 1M tokens)}
\label{tab:llm-pricing}
\end{table}

Despite these limitations, ARISE maintains a modular, auditable, and reproducible architecture. Future work may address these constraints through lightweight model fine-tuning, asynchronous feedback loops with human-in-the-loop reviewers, or more cost-efficient batching strategies.

\subsection*{Broader Impact}
\label{sec:broaderimpact}

ARISE aims to improve the scalability, structure, and factual consistency of academic survey writing by automating key tasks such as citation preparation, outline construction, and LaTeX formatting through modular agent workflows. Its intended users include researchers, educators, and academic writers seeking assistance in synthesizing large volumes of literature.

Beyond survey writing, ARISE’s modular architecture and role-specialized agents can be readily adapted for broader document generation tasks, including grant proposals, technical reports, and educational materials. The rubric-guided feedback loop and structured refinement process generalize well to any domain requiring high-fidelity, structured writing.

The broader impact of ARISE is twofold. Positively, it lowers the barrier to entry for producing well-organized, citation-grounded scholarly outputs. This is especially beneficial in under-resourced research communities or interdisciplinary areas where manual literature synthesis is prohibitively time-consuming. Furthermore, ARISE emphasizes traceable references, rubric-based evaluation, and workflow transparency, promoting responsible deployment and downstream auditing.

However, certain risks must be considered. Over-reliance on automated systems may diminish critical thinking or perpetuate biases inherent in training data. If deployed without expert oversight, ARISE could contribute to derivative content or fail to capture diverse perspectives. To mitigate these risks, ARISE is explicitly designed as an assistive tool—never a replacement for human authorship. All final outputs must be reviewed and approved by domain experts.

We encourage future research to explore participatory reviewer integration, adaptive learning mechanisms, and safeguards to ensure originality, diversity, and attribution fidelity.

\subsection*{Ethical Statement}
\label{sec:ethicalstatement}

We used large language models, including ChatGPT, solely to polish the manuscript’s language, including grammar and phrasing. All substantive content, system design, and experimental decisions were authored and verified by the human research team.

ARISE is designed to assist researchers by automating structured writing tasks such as citation collection, summarization, and LaTeX formatting. It operates on verified academic inputs and uses low-temperature generation to minimize hallucinations. Most observed failure cases stem from external API disruptions (e.g., quota exhaustion) rather than model unpredictability. ARISE is not intended to replace human authorship or scholarly judgment. It produces draft materials for human review, and all final responsibility and intellectual ownership remain with the user.

\section{6. Additional Human Evaluation with GPT-4.1-Mini}

To assess whether our rubric-guided refinement remains effective with smaller models, we conducted an additional human evaluation using GPT-4.1-mini. As before, four expert reviewers (two professors, one postdoc, one PhD student) assessed five survey topics using the same detailed rubric.

As shown in Table~\ref{tab:human_eval_41mini}, the system achieved an average total score improvement from 66.45 to 80.70 (\textbf{+21.45\%}) and a subcategory average increase from 3.32 to 4.03 (\textbf{+21.45\%}). These gains are even higher than the +19.20\% improvement observed with GPT-4.1 (Table~\ref{tab:human_eval}), highlighting that rubric-guided refinement remains robust and impactful even when starting from weaker initial drafts. The smaller model benefits more due to its lower baseline quality, while the full GPT-4.1 starts from a higher base and thus shows slightly smaller relative gains. This comparison underscores the generalizability of our framework across model sizes.

\begin{table*}[t]
\centering
\small
\begin{tabularx}{\textwidth}{lcccc}
\toprule
\textbf{Topic} & \textbf{Before (Total Score)} & \textbf{After (Total Score)} & \textbf{Before (Subcat. Avg)} & \textbf{After (Subcat. Avg)} \\
\midrule
Disease Detection Across Modalities & 69.00 & 82.25 & 3.45 & 4.11 \\
Automated Survey Generation         & 70.00 & 86.25 & 3.50 & 4.31 \\
Retrieval-Augmented Generation      & 65.25 & 79.25 & 3.26 & 3.96 \\
Time-Series Modeling                & 68.00 & 81.00 & 3.40 & 4.05 \\
Topology of Fractal Squares         & 60.00 & 74.75 & 3.00 & 3.74 \\
\midrule
\textbf{Average}                    & \textbf{66.40} & \textbf{80.70} & \textbf{3.32} & \textbf{4.03} \\
\textbf{Improvement (\%)}           &               & \textbf{21.45\%} &               & \textbf{21.45\%} \\
\bottomrule
\end{tabularx}
\caption{Human evaluation results under GPT-4.1-mini. Despite reduced model capacity, the agentic refinement pipeline yields substantial improvements across five diverse topics, as verified by four expert reviewers.}
\label{tab:human_eval_41mini}
\end{table*}

\section{7. Ablation Study: Small-Model Pipeline (GPT-4.1 Mini)} 

To assess the trade-offs between cost and performance, we conducted an ablation study using the GPT-4.1 Mini model throughout the agentic pipeline. As shown in Table~\ref{tab:mini_results}, the smaller model achieved slightly lower initial and final rubric scores compared to the default (full-size) ARISE pipeline, but remained competitive with other systems and published baselines. Notably, we observed that the primary source of performance loss stemmed from the small model's weaker instruction-following and formatting ability (e.g., outputting raw metadata, markdown, or incomplete LaTeX environments). Despite this, the 4.1 Mini pipeline still delivered high-quality survey drafts at significantly reduced computational cost, supporting its use in resource-constrained settings or for rapid iteration.

\begin{table*}[t]
\centering
\small
\begin{tabularx}{\textwidth}{lccc}
\toprule
\textbf{Folder / Topic} & \textbf{Initial Score} & \textbf{Final Score} & \textbf{Avg Improvement per Round} \\
\midrule
Evaluation of Large Language Models               & 85.90 & 91.17 & 1.76 \\
Retrieval-Augmented Generation                    & 79.90 & 90.90 & 11.00 \\
Time-Series Modeling                              & 86.93 & 90.70 & 3.77 \\
Clustering, Indexing and Range Searching          & 88.67 & 90.53 & 1.87 \\
Disease Detection Across Modalities               & 80.10 & 89.67 & 4.78 \\
Integrating Multimodal Fusion, PLMs               & 81.33 & 87.10 & 2.88 \\
Telecommunication                                & 82.50 & 86.60 & 1.03 \\
Generative AI in Mobile Networks                  & 82.50 & 86.60 & 1.03 \\
Generative AI in Manufacturing                    & 82.00 & 86.07 & 1.36 \\
Automated Survey Generation Lit. Review           & 84.43 & 84.87 & 0.22 \\
Topology of Fractal Squares                       & 79.73 & 84.27 & 2.27 \\
\midrule
\textbf{Average}                                 & \textbf{83.09} & \textbf{88.04} & \textbf{2.91} \\
\bottomrule
\end{tabularx}
\caption{Ablation study: Initial and final rubric scores and average improvement per round using the GPT-4.1 Mini model. All scores are averaged across reviewer agents and rubric categories.}
\label{tab:mini_results}
\end{table*}

\bibliography{aaai2026}

\end{document}